\documentclass{article}
\usepackage{hyperref, url}
\usepackage{amscd,amsmath,amsthm,amssymb}
\usepackage{enumerate,varioref}
\usepackage{epsfig}
\usepackage{graphicx}
\usepackage{subfig}
\usepackage{mathtools}
\usepackage{tikz}
\usepackage{algorithmicx}
\usepackage{algorithm}
\usepackage{algpseudocode}
\usepackage{xcolor}

\theoremstyle{definition}

\theoremstyle{remark}

\newtheorem*{rem}{Remark}

\newcommand{\half}{\frac{1}{2}}

\newcommand{\ve}{\vec{1}}

\title{ Picking Efficient Portfolios from 3,171 US Common Stocks with New Quantum and Classical Solvers}
\author{Chicago Quantum\footnote{Jeffrey Cohen, Clark Alexander} \\ email \href{mailto:research@quantum-usaci.com}{the authors}}

\begin{document}

\maketitle

\begin{abstract}
	We analyze 3,171 U.S. common stocks to create an efficient portfolio based on the Chicago Quantum Net Score (CQNS).  We begin with classical solvers, then incorporate quantum annealing.  We incorporate a simulated bifurcator into our `team’ of solvers, along with the new D-Wave Advantage$^{TM}$ quantum annealing computer with 5,760 available qubits in a Pegasus graph size P16 architecture.	
\end{abstract}

\tableofcontents

\section{Introduction}

In this work we continue our progress on the stock picking problem using the Sharpe ratio and the Chicago Quantum Net Score.  Previously we had worked through the problems of picking the optimal portfolio from a universe of 40 then 60 stocks.  In this work we have scaled our problem to a universe of 3171 U.S. common stocks, which in some measure is the entirety of all daily traded stocks which have ``good data" for the year period beginning September 2019 and ending September 2020.  In order to scale our solutions for this problem we have employed a number of new techniques.  We continue to use a genetic algorithm, several simulated annealers, and the D-Wave Systems (hereafter D-Wave) 2000Q quantum annealing computer.  In this work we employ D-Wave's next generation quantum annealing computer, the Advantage$^{TM}$ \cite{D-WaveAdvantageOverview}, as well as our own simulated bifurcation machine.  We have significantly sped up our genetic algorithm to allow for many more generations and to produce an initial population from multiple sources.  
While we have worked through a universe of 3171 stocks, we have no claim to a universal best solution, but rather, we claim to have gotten ``good" solutions, which we can measure empirically by their performances as stock portfolios.  Since we are incorporating a buy-and-hold strategy, we do not consider puts and calls, or the Black-Scholes model with the American or European option, nor do we employ the plethora of time series based analyses of stocks. Instead, we present our results from the work we have done. This set of techniques leaves us with a solution space of $2^{3171}$ possible portfolios.  A quick calculation reveals:
\[
2^{3171} \approx 3.68 \times 10^{954} 
\]
Being late 2020, we can precisely say what computationally intractable means.  In this case, let's assume we have exascale (digital computers) that is, $10^{18}$ FLOPS.  Additionally, if every atom in the observable universe were its own exascale computer we'd arrive at approximately $10^{90}$ exascale computers, and we have had roughly $4.4 \times 10^{17}$ seconds.  Thus we round up to $10^{18}$.  Putting it toghter, there is no way to brute force a solution of size $10^{126}$.  We see that we are far in excess of this number with 3171 stocks.  Thus while we know that our solutions are better than quintillions of other possible portfolios, we cannot in good faith guarantee a universal optimal solution.  However, one faces a second issue in measuring the ``goodness" of a solution when dealing with stocks.  Is the investor doing better than the market? If so, how much better?  In this case, our solutions will, with high probability, pass muster.

\section{Motivation}

Our work begins with validating and running all U.S. equities (common stocks) through our classical solver to find the most attractive portfolio of $N$ stocks that can be run on a quantum annealing computer.  We also allow these solvers to identify attractive portfolios.  In Step 2, we take the best $N$ stock portfolio found and run those stocks through additional solvers, including the quantum annealing computer, to identify the overall most attractive portfolios.

This two-step approach enables us to more efficiently find deeper local minima i.e. more negative CQNS values, which imply better investment portfolios.  Our two-step process finds lower CQNS scores, which equate with more attractive portfolios. 

Our motivation is that our classical solvers scale to analyze U.S. common stocks listed on NYSE, NASDAQ-Q and NYSE American (previously the American Stock Exchange), which includes 3,171 validated stocks as of October 10, 2020.  At this scale, we create a single $3171\times 3171$ matrix for each portfolio size desired to create our QUBO.

The D-Wave quantum annealing computer scales to analyze 138 out of a maximum possible 180 \cite{MaxCliques, 180Cliques} stocks at one time (20\% embedding success rate) and embeds 134 stocks consistently ($\sim$80\% embedding success during this experiment).  We have additional classical solvers at this scale, including the D-Wave TABU and simulated annealer, along with our in-house simulated bifurcation machine.  These methods utilize the QUBO which we create at this scale.

For 3,171 stocks, we use the following workflow.
\begin{enumerate}
	\item Calculate the `all-in’ values, and calibrate CQNS value to zero.
    \item Run Monte Carlo in a discrete probability distribution around $N/2$ stock portfolio size
    \item Run Monte Carlo for target $N$ portfolio size
    \item Run Genetic Algorithm, with a randomly selected initial population for target portfolio size of $N$.
    \item Run the simulated annealer for target portfolio size of $N$.
    \item Build $N$-asset, $M\times M$ (here $M=3171$) matrix for conversion to QUBO
    \item Run D-Wave simulated annealer for target portfolio size of $N$ 
    \item Run the simulated bifurcation machine for target portfolio size of $N$
    \item Run D-Wave TABU sampler for target portfolio size of $N$
    \item Run the Genetic Algorithm with prior solution as its initial population,  for target portfolio size of $N$.
\end{enumerate}

We adjust the parameters for each solver to run each of them in five minutes or less.  In our experiment we did not find a valid solution (correct number of assets) with the D-Wave TABU or Simulated Annealer samplers.  They return larger solutions (1,500 and 350 vs. 134 stocks).

Our matrix build creates the QUBO data for each size stock portfolio in sequence, and adds the values to the $N\times N\times N$ array.  Our matrix build fails between 1,500 and 2,000 matrices, likely due to a memory issue with our iMAC 2013.  This occurred whether we broke up the build process into subsets, or we eliminate the matrix scale process.

In the end, our best $N$, or 134 stock portfolio, had a CQNS score of $-3.14\times 10^{-3}$.
This was found by our seeded genetic algorithm.  The `all 3,171 asset’ portfolio has a CQNS score of zero, and the other solvers found solutions between these values if they found valid solutions. 

\begin{rem} Our collective runs took 232 seconds. This gives a bit of scale as to how fast the SBM runs on home scale architecture.  We can speed this up significantly on a more powerful computing platform.  
\end{rem}


\section{The Simulated Bifurcation Machine}

Our work with the simulated bifurcation machine, hereafter SBM, is based fully on \cite{SB}.  We shall not go into the explanation of why the mechanism of SBM works, but rather we shall say what modifications we made, why, and what results we obtained using SBM.  One should also note, that we did not use a version of SBM available through a cloud service, but rather one that was coded in-house, the reason for this is simple, we want to really understand how to tune the parameters to force the bifurcations to happen as quickly or as slowly as desired.

\begin{figure}[!ht]
	\centering
	\includegraphics[width=0.75\textwidth]{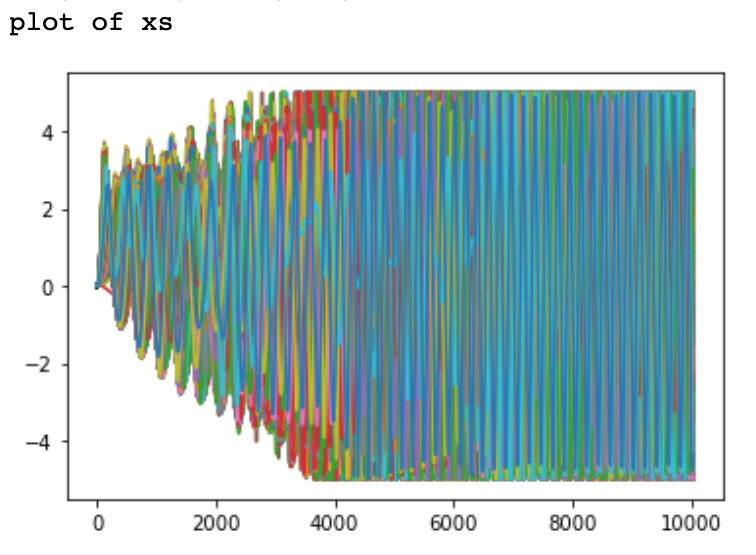}
	\caption{Time evolution of $x$ vector in Simulated Bifurcation}
	\label{fig: SBM xs}
\end{figure}

In \ref{fig: SBM xs} we see real data, but we acknowledge that 3171 items is difficult to visualize, so we have added some smaller portfolios in Appendix B \ref{fig: additional SBM}.

\subsection{Our Modifications}
The most important modification we should mention is that the official white paper contains a small error for our set up.  The set up of SBM in the white paper assumes a purely Ising model matrix for a QUBO model.  In our case, we are looking at a $\{0,1\}$ solution set rather than a $\{-1,1\}$ solution set.  This means we must make a coordinate change.  

\begin{rem}
	Let $\vec{x} \in \{0,1\}^N$ and $\vec{z} \in \{-1,1\}^N$.  One converts $\vec{x}$ to $\vec{z}$ by
	\[
	z_i = 2x_i - 1.
	\]
	Thus we define $\ve = \sum \hat{e_i}$ and write
	\[
	\vec{z} = 2\vec{x} - \ve
	\]
	
	A generalized quadratic form is given by
	\[
	\vec{x}^tA\vec{x} + B\cdot\vec{x} + K_1
	\]
	We wish to convert this to an Ising model 
	\[
	\vec{z}^t J \vec{z} + C\cdot\vec{z} + K_2
	\]
	
	To make a long story short we get the following formulae
	\begin{eqnarray}
	J & = & A/4\\
	C & = & \half \left( B + 4(J\ve)\right)\\
	K_2 & = & K_1 + 2(C\cdot\ve) - \ve^tJ\ve
	\end{eqnarray}
	
	However, we throw away both $K_1$ and $K_2$ since they are unimportant to the values of $\vec{x},\vec{z}$.
	
\end{rem}

The main modifcation we make is that in converting from a pure $\{0,1\}$ quadratic form i.e. $\vec{x}^t A \vec{x}$ to an Ising model, we pick up a nonzero offset vector $C$.  In \cite{SB} this offset vector is missing in equations 12,13,17, and thus in our coding we have reintroduced these offset vectors.

That is to say we have made the change
\begin{equation}
\xi_0 J\cdot\vec{z} \mapsto \xi_0( J\cdot\vec{z} + C)
\end{equation}

We also experimented with the pressure scaling from a linear progression of zero to one to a logistic curve, however we found no empirical difference in the quality of solutions.  

The other modifcation we have made is that when one trajectory has clearly bifurcated, but others are still oscillating around zero, we have set a threshold to avoid divergent trajectories.  We were unable to replicate the quality of solutions in the original paper for large scale problems, however in tiny problems, (i.e. dozens of trajectories) we were able to find the universally best solution.  We did, however, manage to tune our SBM to give good solutions very quickly.  Our empirical findings, however, show that for a single solution technique, our in house simulated annealer tuned with the statistics of random matrices was able to consistently produce better results, at the cost of additional time.  For example, in time tests, we were able to produce very good results on a $10,000\times 10,000$ matrix at the cost of 47 minutes, but on a laptop.  For this paper we executed 10,016 iterations on a $3,171\times 3,171$ matrix to select 134 stocks in 232 seconds.  It is, obviously the reader's choice as to which combination of techniques to use, but we find that the combination of seeding a genetic algorithm with a initial population of Monte Carlo, simulated annealer, genetic algorithm, quantum annealer, and SBM solutions works best.

\subsection{Tuning the SBM}

The simulated bifurcation machine takes in a number of parameters for tuning, specifically $\Delta, K, \xi, \varepsilon, p$, and number of iterations.  In our experiments we found that the most effective parameters for tuning were $\varepsilon, \xi,$ and number of iterations.  The white paper \cite{SB} sets $\Delta, K$ at approximately 1, and $p$ increases linearly.  We found these parameter choices to be sufficient, but the coupling constant $\xi$ which drives the adiabatic shift from nonlinear oscillator to Ising model greatly affected our results.  Additionally, $\varepsilon$ affected how quickly a bifurcation occurred, and number of iterations is roughly inversely proportional to $\varepsilon$.

\subsection{Our Picks}

The stocks we selected out of 3,171 candidates were AX and MCRB, with data effective from market close October 9, 2020.  This analysis is not fundamental to the companies, but relies on the patterns of the adjusted closing price data.  

Axos Financial, Inc. (AX) is a regional bank with mortgage exposure in Southern California.  
Seres Therapeutics, Inc. (MCRB) is a microbiome therapeutics biological drug platform company. These are picks from significantly different industries.  

They started up 5\% and 0.5\% in the first 2 and a quarter trading days, compared to the market indices which are up $\sim 1$\%.  Our model also appears to select some stocks with short-term volatility to trade on due to preferring stocks with higher $\beta$ values.

\begin{figure}[!ht]
	\centering
	\includegraphics[width=0.75\textwidth]{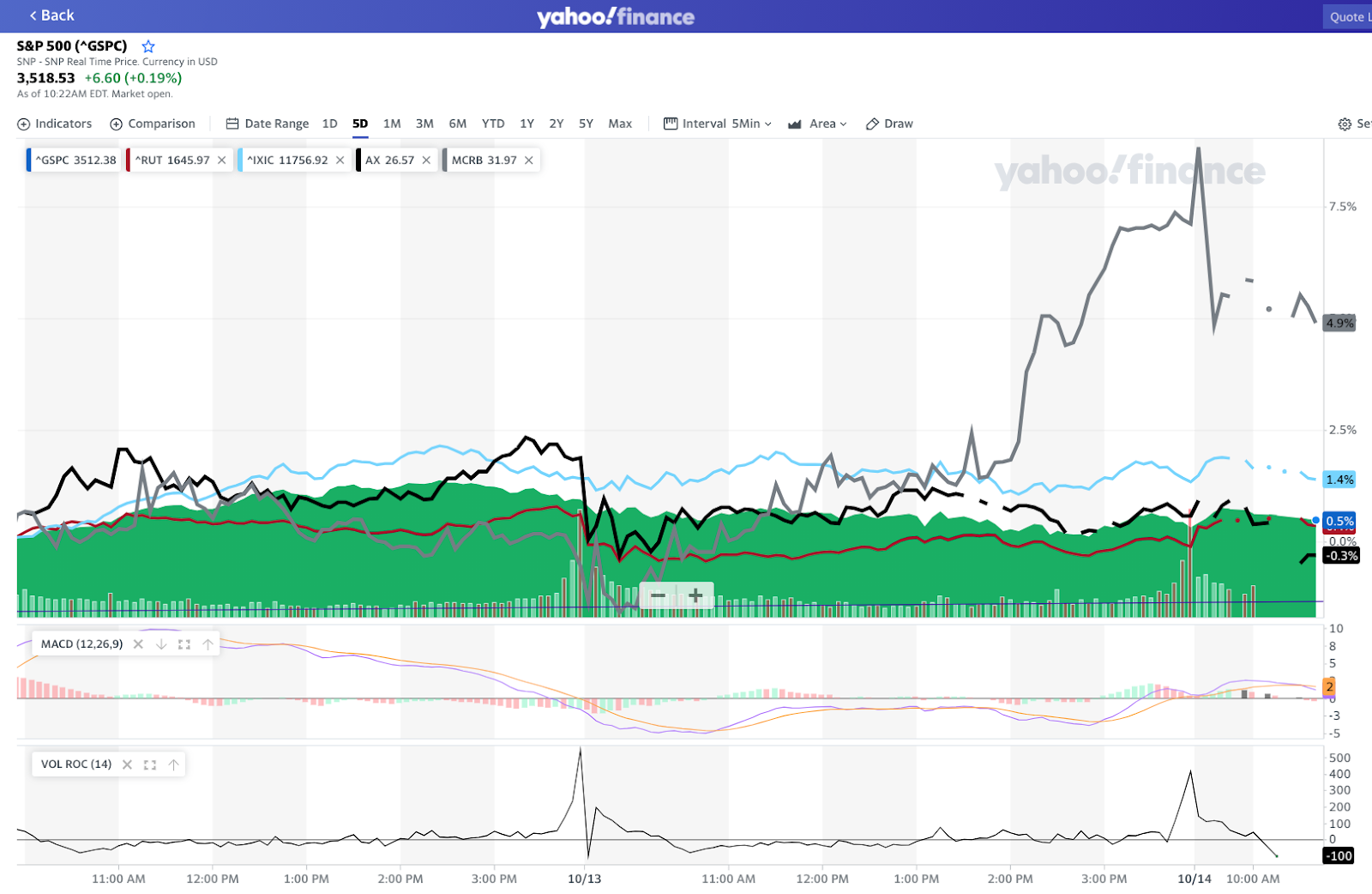}
	\caption{Movement of the pair over 2.25 trading days}
	\label{fig:stockpicks}
\end{figure}

\begin{figure}[!ht]
	\centering
	\includegraphics[width=0.75\textwidth]{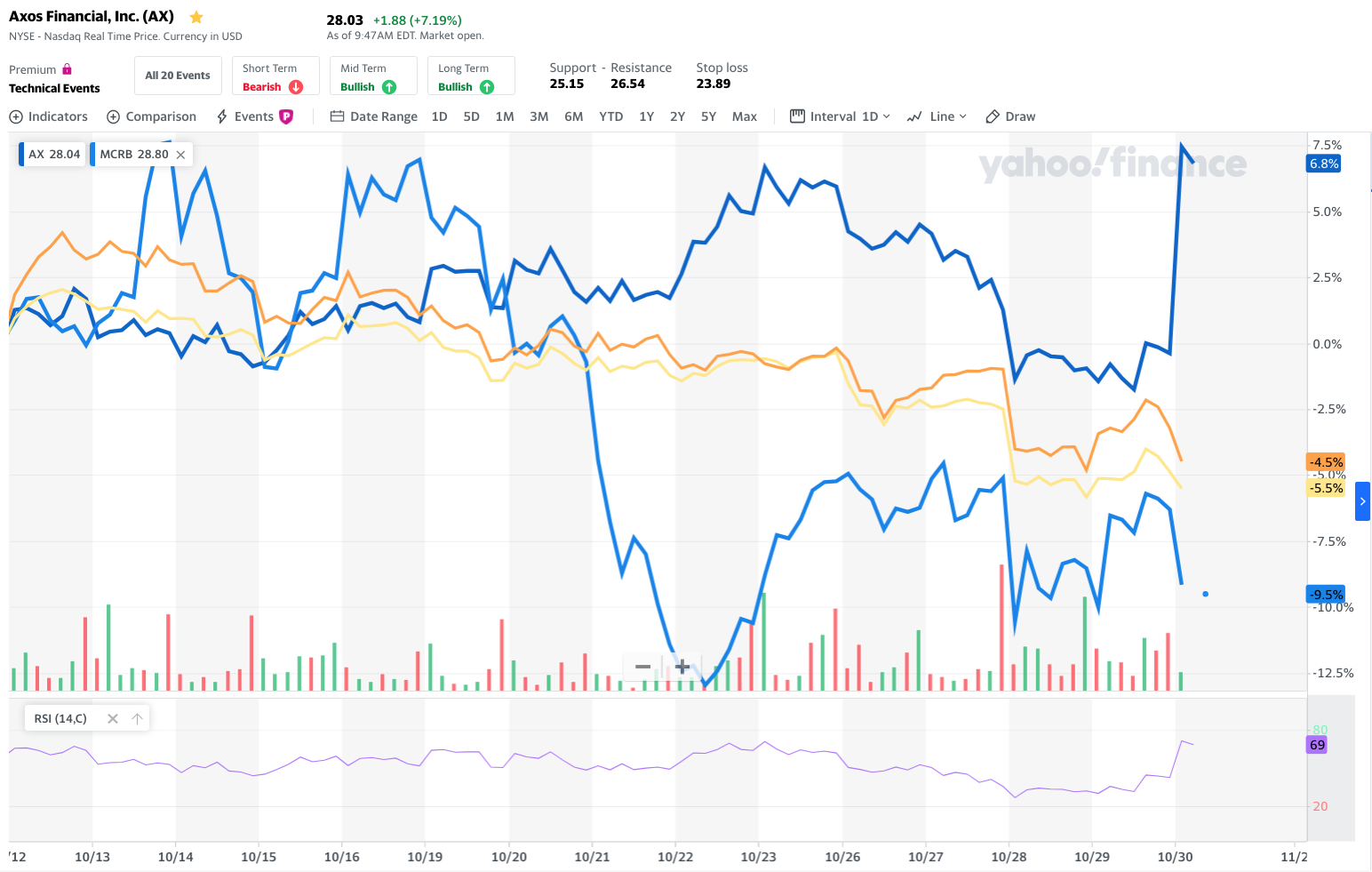}
	\caption{Movement of the pair over 14.5 trading days, orange is QQQ, yellow is SPY, dark blue is AX, lighter blue is MCRB}
	\label{fig:stockpicks2}
\end{figure}

We check the performance of our two stocks against the S\&P 500 index ETF (SPY) and the index we use to calculate $\beta$. We start October 12 market open and we read out October 30 at time of writing.  (MCRB) and (AX) appear as blue lines, while the index is yellow.

While the S\&P 500 fell 5.5\%, (AX) increased 6.8\%, and (MCRB) fell 9.5\%.  On average, our equally weighted portfolio falls 1.4\% which is less than half the (SPY). \ref{fig:stockpicks2}

These stocks consistently move in opposite directions over the period and straddle the index.  (MCRB) has significant volatility during the period which gives investors a chance to trade that stock.

In this chart, the Nasdaq Composite index ETF (QQQ) fell 4.5\% during the period. The (QQQ) also moves between the two stocks selected. 

\section{Scaling the D-Wave Advantage Solver}

The new D-Wave computer has 5,760 qubits set in a ``Pegasus" graph size P16 topology with increased connectivity over the ``Chimera" topology in the previous solver generation.  It runs at $15.8 mK$, and is used via D-Wave’s Ocean$^{TM}$ software development toolkit available through the D-Wave Leap$^{TM}$ cloud-based service.

\begin{figure}[!ht]
	\centering
	\subfloat{{\includegraphics[width=5.5cm]{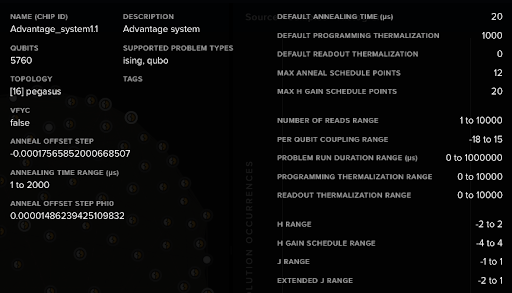}}}
    \subfloat{{\includegraphics[width=5.5cm]{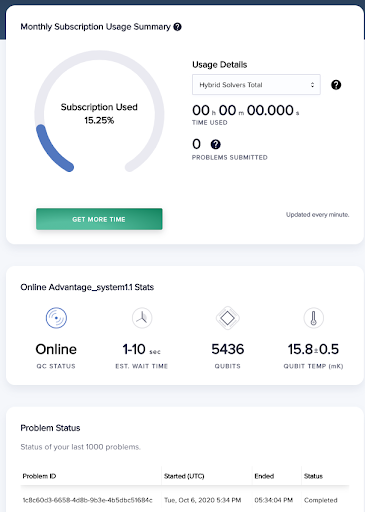}}}
    \caption{D-Wave Dashboard}
\end{figure}

From an experimental perspective the D-Wave Advantage 1.1 runs like the previous generation of quantum annealing solver. We connect via a python Jupyter Notebook, establish the connection via the cloud, and submit jobs for execution.  We use the D-Wave Problem Inspector$^{TM}$ to gain additional information about the problem submitted, its embedding on the qubits, and the system itself.

We successfully embedded 138 stocks on the system and downsize the problem to 134 stocks for our experiment to increase embedding success rates.  

\subsection{Changes from the Previous Generation}

\begin{figure}[!ht]
	\centering
	\subfloat{{\includegraphics[width=5.5cm]{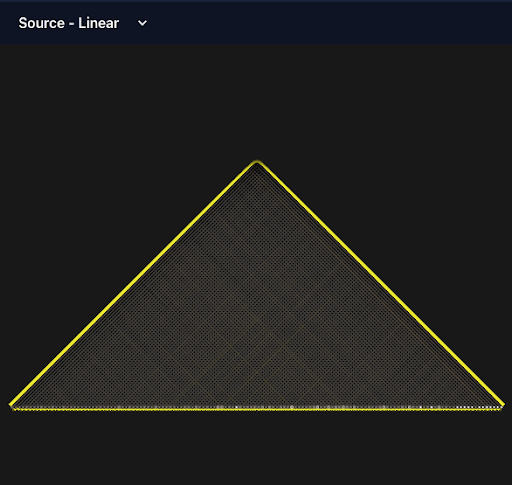}}}
	\subfloat{{\includegraphics[width=5.5cm]{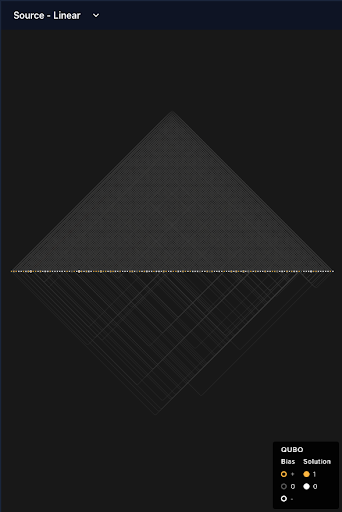}}}
	\newline
	\subfloat{{\includegraphics[width=5.5cm]{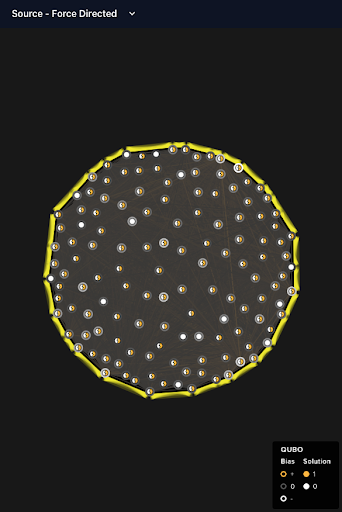}}}
	\caption{Chain Strengths and Connections}
	\label{fig: Chain Strengths}
\end{figure}

A few things change during this experiment from our prior work on the D-Wave Chimera system (embed 64 stocks).  
\begin{enumerate}
\item The wait time between runs has increased significantly to $\sim 10$ minutes  on average depending on the time of day.  This requires us to significantly limit our runs.
\item D-Wave embedding was successful 11 of 15 attempts for 134 stocks.  We embedded and ran a maximum of 138 stocks one time.  When we fail to embed an error occurs and we manually restart the analysis.
\item We see biases upwards of (-7,3) on the qubits, whereas in the Chimera our bias range was typically (-1,1).  
\item Qubit bias ranges of (-1,1) create qubit connections above and below the matrix.  Qubit biases that significantly exceed abs(1.0) due to division by chain strength values create no connectivity below the matrix (see 2 charts below).\ref{fig: Chain Strengths}
\item The chain strengths that find valid solutions are 0.15 and 1.5.  
\item Our chain lengths vary from 27 to 45 qubits. Our experimental results show almost 100\% chain breaks which is significantly higher than on the Chimera.
\item The number of target variables, and max chain length, varied due to the embedding quality.  Best and worst combination:\\ 

\begin{tabular}{c | c | c }
	target variables & max chain length & chain strength\\
	\hline
	2,430  & 27  & 0.2\\
	3,730  & 45  & 1.5
\end{tabular}\\

\item We scale our QUBO to (-4,4) instead of (-0.99,0.99) before using it with solvers.
\item We vary the number of runs for each size portfolio from 200 to 1000.
\item We vary the annealing times from 20 to 150 $\mu$s with the portfolio size desired (larger portfolios require more annealing time).
\item As for all methods, we save all ``better" CQNS portfolios from each method for use as input to the seeded genetic algorithm. 
\item Our change in formulation, for a portfolio $P$, from  $Var[P] - \Sigma_i(\mathbb{E}[P_i]^3)$ to $Var[P] - \left(\Sigma_i \mathbb{E}[P_i]\right)^3$ for the final portfolio scoring does change the resulting ``ideal" portfolio.  We are using the QUBO-based solvers to select portfolios based on the previous formulation.  We changed this to enhance the quality of our portfolio selection.
\item The QPU run time on the D-Wave systems varies from 85k to 199k $\mu$s with programming times consistently at 26k $\mu$s and annealing times set by the operator.  Post processing times vary from 8k to 36k $\mu$s.
\item The number of stocks we seek in the QUBO matters.  We sampled for 5 portfolio sizes, were successful with 67, 47 or 3 stocks out of 134, but not 37 or 17 stocks.  We envision running through more portfolio sizes once the system wait time is reduced and the embedding success rate increases.
\end{enumerate}

\subsection{Experimental Results}

We controlled for sample size between 3,171 stocks and 134 stocks by setting the CQNS\_power = $\ln(Var) / \ln(\mathbb{E})$.  We then used the same CQNS\_power with 134 stocks.  Our first portfolio chosen, with 134 stocks, had a CQNS score $-3.14\times 10^{-3}$, versus the second portfolio chosen, with 2 stocks, which had a value of $-3.9\times 10^{-3}$.  The best answer was found by four of our solvers (GA seeded, MC, bespoke SA, and DWave SA).  

The best D-Wave quantum annealing run returned 3 stocks which was less attractive than our ideal solution.  It had a CQNS score of $-1.69\times 10^{-3}$, was found in 234k $\mu$s plus wait and cloud access time.  This is a good solution, better than all-in 134 stocks with a value of $-6\times 10^{-5}$ and the simulated bifurcator score of $4.1\times 10^{-4}$.

\section*{Appendix A: Preparing the Data}

We FTP the NYSE and NASDAQ Q stocks from \cite{NASDAQ} FTP site, then capture the NYSEAmerican stocks (Tape B)\cite{NYSE} through a manual download (in xls format) from the NYSE website.  We remove funds, trusts, warrants, test, and preferred stocks. Our stock sample includes US inter-exchange listed foreign stocks that trade in U.S. exchanges.

We do a test download of each stock for one year and remove stocks with negative prices, missing prices, failed downloads, and less than one full year of continuous trading data via the yfinance python module.

We check for positive eigenvalues (within precision tolerance of python) in the covariance matrix created with all stocks to ensure we have a definite positive covariance matrix.

We set a $\beta$ range and remove stocks with higher and lower $\beta$ values.  In this experiment, we only remove negative $\beta$ stocks.

\section*{Appendix B: Charts}

\begin{figure}[!ht]
	\centering
	\subfloat[CQNS]{{\includegraphics[width=5.5cm]{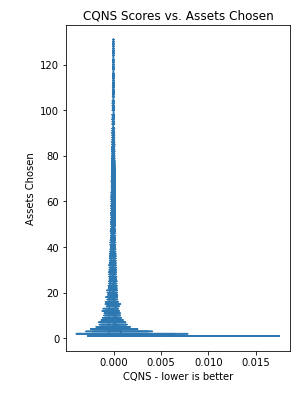} }}%
	\qquad
	\subfloat[CQNS vs. Portfolio Sequence]{{\includegraphics[width=5.5cm]{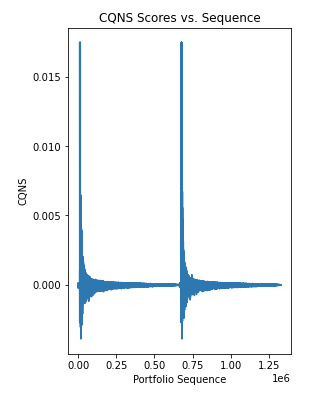} }}%
	\newline
	\subfloat[CQR vs. Portfolio Sequence]{{\includegraphics[width=5.5cm]{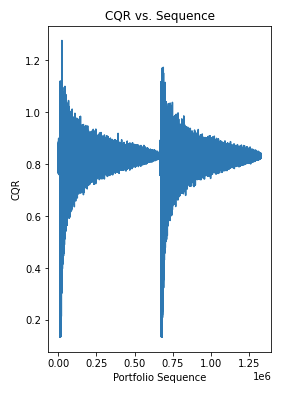} }}%
	\qquad
	\subfloat[Sharpe Ratio vs. Portfolio Sequence]{{\includegraphics[width=5.5cm]{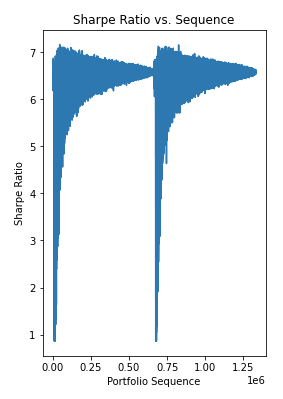} }}%
	\caption{Charts showing resulting scores with different methods
	}
	\label{fig:CQNS charts}
\end{figure}

\begin{figure}[!ht]
	\centering
	\includegraphics[width=0.75\textwidth]{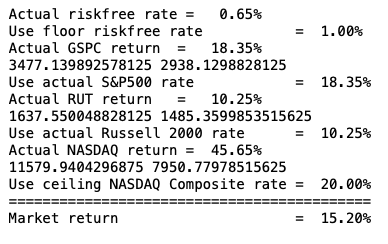}
	\caption{A time of rising markets}
	\label{fig:Rising Markets}
\end{figure}

\begin{figure}[!ht]
	\centering
	\subfloat[Qubit Embedding 1]{{\includegraphics[width=5.5cm]{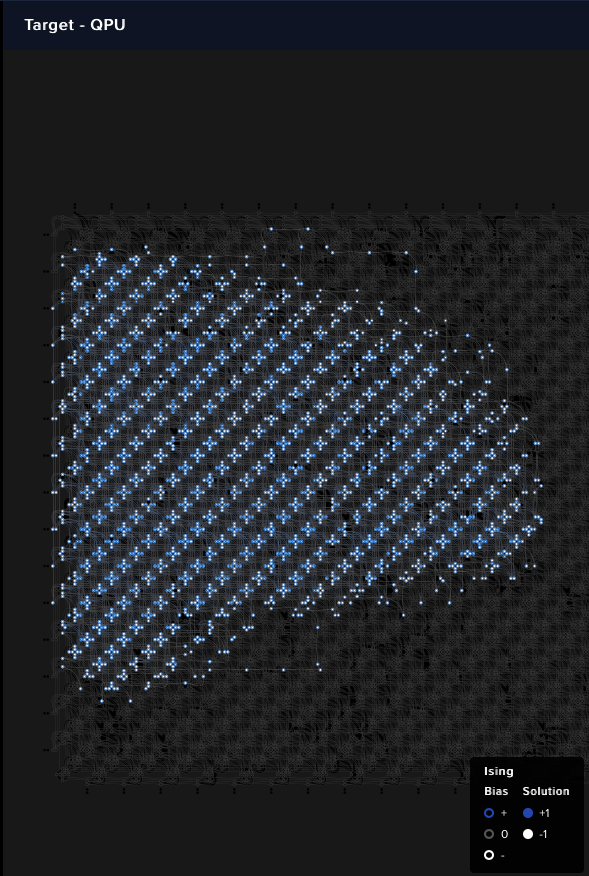} }}%
	\qquad
	\subfloat[Qubit Embedding 2]{{\includegraphics[width=5.5cm]{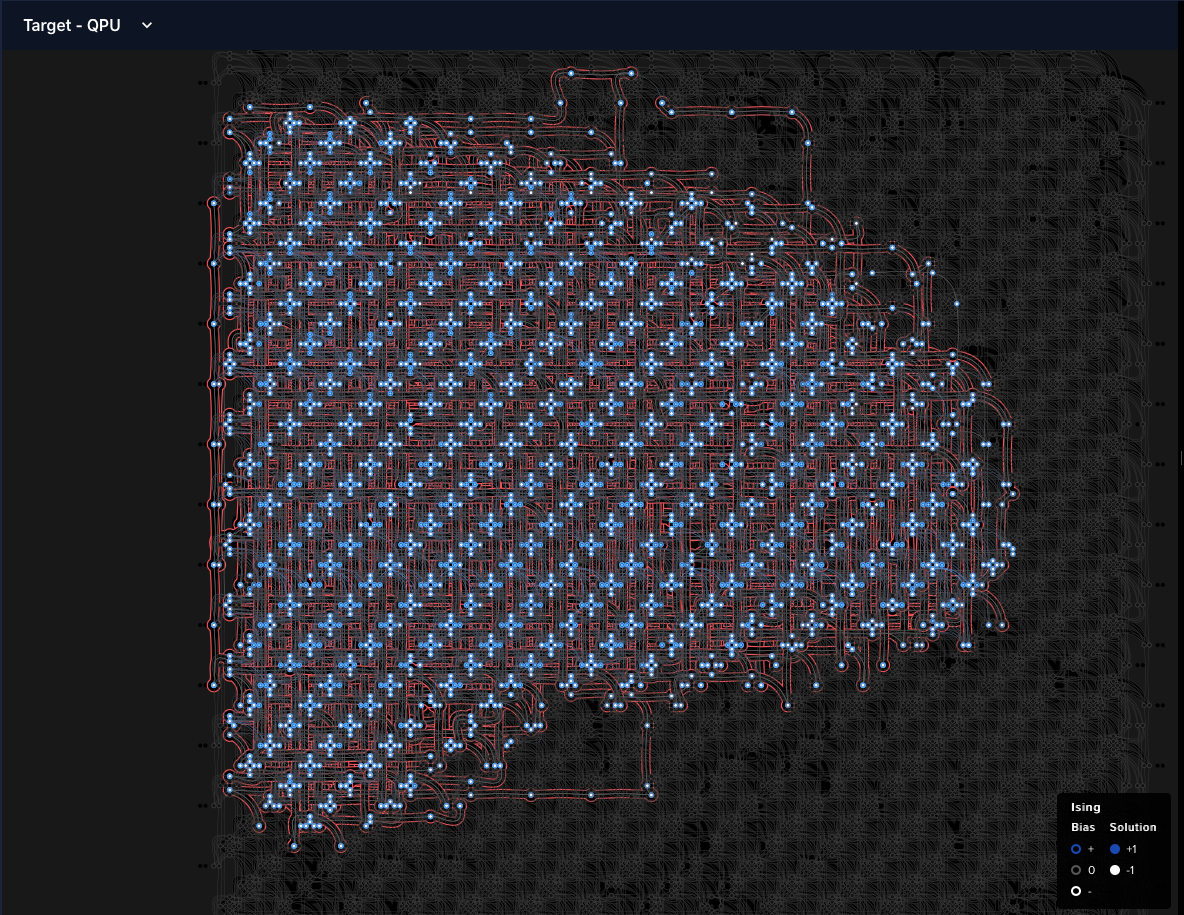} }}%
	\newline
	\qquad
	\subfloat[Qubit Embedding 3]{{\includegraphics[width=5.5cm]{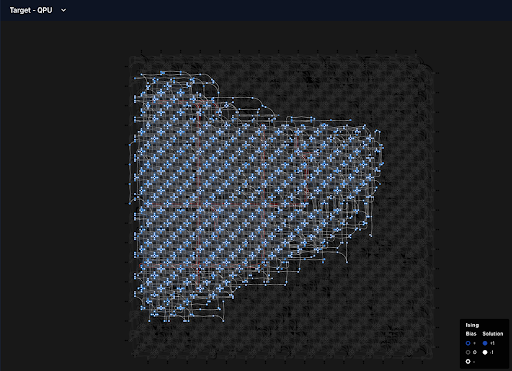} }}%
	\qquad
	\caption{Charts showing different qubit embeddings}
	\label{fig:CQNS charts 2}
\end{figure}

\begin{figure}[!ht]
	\centering
	\subfloat[A 25 stock run of the SBM]{{\includegraphics[width=5.5cm]{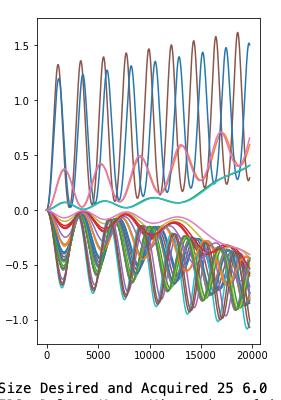}}}
	\qquad
	\subfloat[A 15 stock run of the SBM]{{\includegraphics[width=5.5cm]{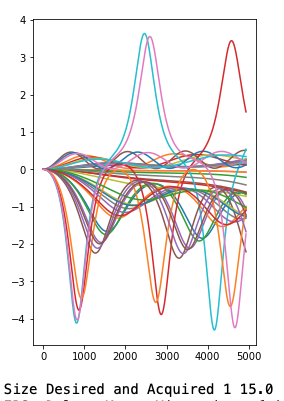}}}
	\qquad
	\caption{Additional runs of Simulated Bifurcator Machines}
	\label{fig: additional SBM}
\end{figure}

\end{document}